# Observation of Valley-polarized Landau Levels in Strained Graphene

Si-Yu Li[1,2], Ke-Ke Bai[1,2], Long-Jing Yin[1,2], Jia-Bin Qiao[1,2], Wen-Xiao Wang[1,2], Lin He[1,2],*

[1] Department of Physics, Beijing Normal University, Beijing, 100875, People's Republic of China.

[2] The Center of Advanced Quantum Studies, Beijing Normal University, Beijing, 100875, People's Republic of China.

*Correspondence to: helin@bnu.edu.cn.

**Abstract**: In strained graphene, lattice deformation can create pseudo-magnetic fields and result in zero-field Landau level-like quantization. In the presence of an external magnetic field, valley-polarized Landau levels are predicted to be observed because the pseudo-magnetic fields are of opposite directions in the *K* and *K'* valleys of graphene. Here, we present experimental spectroscopic measurements by scanning tunneling microscopy of strained graphene on Rh foil. We direct observed valley splitting of the Landau level induced by the coexistence of the pseudo-magnetic fields and external magnetic fields. The observed result paves the way to exploit novel electronic properties in graphene through the combination of the pseudo-magnetic fields and the external magnetic fields.

**One Sentence Summary:** Valley-polarized Landau quantization is realized in strained graphene via combination of pseudo-magnetic fields and external magnetic fields.

**Main Text:** The linear band dispersion of graphene allows the existence of a variety of mechanisms which generate effective gauge fields in the low-energy electronic theory (*1-4*). Very recently, it has been demonstrated explicitly that a spatially varying lattice distortion in graphene gives rise to pseudo-magnetic fields acting on the low-energy charge carriers and results in Landau level-like quantization even in the absence of a magnetic field (*5-7*). The primary difference between the strain-induced pseudo-magnetic field $B_S$ and the external magnetic field $B$ is that the $B_S$ preserves time-reversal symmetry and has opposite signs in the two low-energy valleys, $K$ and $K'$, of graphene (*1*, *8-12*). Therefore, the combination of the pseudo-magnetic field and the magnetic field could lift the valley degeneracy of the Landau levels (LLs) and lead to unconventional valley-polarized Landau quantization in graphene monolayer.

Here, we report the experimental measurement of the valley-polarized LLs in strained graphene on Rh foil. The graphene monolayer was grown on Rh foil via a traditional ambient pressure chemical vapor deposition (CVD) method (*13*). Because of mismatch of thermal expansion coefficients between graphene and the supporting substrates, strained structures could easily be observed for graphene grown on metallic substrates (*5-7*, *14*, *15*), see Fig. S1. Figure 1A and 1B show representative scanning tunneling microscopy (STM) images of a strained graphene structure. Such one-dimensional quasi-periodic graphene ripples may arise from the anisotropic surface stress of the substrate (*14*, *15*).

Asymmetry of local density of states (DOS) of electrons and holes has been observed previously in strained graphene (*5*, *7*). In our experiment, we direct measure the electron-hole asymmetry in the strained graphene quantitatively through Landau quantization, as shown in Fig. 1C. The scanning tunneling spectroscopy (STS) spectra (Fig. 1C) recorded in the strained region under magnetic fields exhibit Landau quantization of massless Dirac fermions in graphene monolayer, with its characteristic non-equally-spaced energy-level spectrum of LLs and the hallmark zero-energy state (*16-18*). This is consistent well with our previous result that the graphene sheet on Rh foil behaves as pristine graphene monolayer (*13*). Obviously, the peaks of LLs for filled-state are much more pronounced than that for empty-state, indicating a large electron-hole asymmetry in the strained graphene region. The observed LL energies $E_n$ depend on the square-root of both level index $n$ and magnetic field $B$:

$$E_n = \text{sgn}(n)\sqrt{2e\hbar v_F^2 |n| B} + E_0, \qquad n = ...-2,\ -1,\ 0,\ 1,\ 2... \qquad (1)$$

Here $E_0$ is the energy of Dirac point, $e$ is the electron charge, $\hbar$ is the Planck's constant, $v_F$ is the Fermi velocity, and $n > 0$ corresponds to empty-state (holes) and $n < 0$ to filled-state (electrons) (*16-18*). According to Eq. (1), we can obtain the Fermi velocities of electrons $v_F^e$ and holes $v_F^h$ separately. The linear fit of the experimental data to Eq. (1), as shown in Fig. 1D, yields $v_F^e = (1.257 \pm 0.009) \times 10^6$ m/s and $v_F^h = (0.930 \pm 0.014) \times 10^6$ m/s. Such a large electron-hole asymmetry is mainly attributed to the enhanced next-nearest-neighbor hopping $t'$ by lattice deformation and curvature in the strained graphene region (*5, 7, 19, 20*). Theoretically, a nonzero $t'$ in graphene could increase the Fermi velocity of the filled-state, whereas decrease the Fermi velocity of the empty-state (*19, 20*). A finite value of $t' \sim 0.25t$ accounts well for the observed electron-hole asymmetry of the Fermi velocity (here $t$ is the nearest-neighbor hopping parameter, see Supplementary Materials for further discussion).

Besides the electron-hole asymmetry, we observe two other notable features in the STS spectra of some strained graphene regions, as shown in Fig. 2. One feature is the emergence of several peaks: a pronounced peak at the charge neutrality point $N_C$ and several weak peaks at relatively high bias, in the spectra even measured in the absence of magnetic field, see Fig. 2A and Fig. S2. The other feature is the splitting of $n = -1$ LL of the spectra under different magnetic fields, see Fig. 2A and Fig. S3. For the spectra recorded at the same position, the energy difference of the two peaks of $n = -1$ LL, $\Delta_{-1}$, decreases with increasing the magnetic fields (Fig. 2A). For the spectra recorded at the same magnetic field, 4 T, the value of $\Delta_{-1}$ changes over different positions (Fig. 2B).

These experimental observations can be understood within a theoretical framework that incorporates the effects of both strain-induced pseudo-magnetic fields and external magnetic fields on the Landau quantization of massless Dirac fermions in graphene monolayer. For the case of $B = 0$ T, the pronounced peak at the $N_C$, as shown in Fig. 2A, is attributed to the strain-induced partially flat bands ($n = 0$ LL) at zero energy (*7*, *21*, *22*), and the weak peaks at high bias are attributed to higher pseudo-Landau levels. The signal of the higher pseudo-Landau levels is rather weak, which is in agreement with earlier STM measurement in strained graphene ripples (*7*). This result is also quite reasonable because the description of a hopping modulation in strained graphene as an effective pseudo-magnetic field is exact valid only at the $N_C$ and the higher pseudo-Landau levels are less well defined (*21*, *22*). From the energy spacing between the $n = 0$ LL and the $n = -1$ LL measured in $B = 0$ T, the pseudo-magnetic field is estimated to be about $(0.61 \pm 0.05)$ T according to Eq. (1) (In the calculation of $B_S$, we use $v_F^e = (1.250 \pm 0.007) \times 10^6$ m/s obtained in our experiment). Theoretically, pseudo-magnetic field arises from spatially variation of nearest-neighbor hopping $t$ in graphene (see Supplementary Materials for further discussion) (*1*,*2*). The result of Fig. 2 indicates that the local strain in the region where the spectra of Fig. 2 are recorded not only enhances the $t'$, but also results in spatially variation of $t$. A local strain of 1% in graphene is predicted to generate the pseudo-magnetic field of 10 T (*22*). Therefore, the observed $B_S \sim 0.6$ T is quite reasonable with considering the width and height of the studied nanoscale ripples.

In the presence of both the pseudo-magnetic field and the external magnetic field, the total effective magnetic field in one of the valleys, for example the K valley, is $B - B_S$, and that of the other valley, for example the K' valley, is $B + B_S$, as schematically shown in Fig. 3A. Then, the valley degeneracy of the $n \neq 0$ LLs is expected to be lifted (*9,12*) and the energy spacing of the two valleys for the $n_{th}$ LL should be

$$\Delta_n = \sqrt{2e\hbar v_F^2 |n|(B+B_S)} - \sqrt{2e\hbar v_F^2 |n|(B-B_S)}, \quad n = ...-2, -1, 1, 2... \quad (2)$$

With considering the fact that the picture of the effective pseudo-magnetic field is less well defined at high energy, the valley-polarized LLs should be observed only for small Landau indices $|n|$, for example for $n = \pm 1$ LLs. In our experiment, the valley-polarized LL is only observed for $n = -1$, as shown in Fig. 2A. Although the absence of splitting of $n = 1$ LL measured in $B \neq 0$ T remains to be understood, it is likely that these observations are closely linked to the electron-hole asymmetry observed in our experiment. Figure 3B summarizes the values of $\Delta_{-1}$ as a function of the external magnetic fields recorded at the same position of the strained graphene. The fit of the experimental result to Eq. (2) yields $B_S = (0.59 \pm 0.03)$ T, which is consistent well with that estimated from the energy spacing of pseudo-Landau levels measured in zero magnetic field. The dependence of the values of $\Delta_{-1}$ recorded at 4 T on different positions, as shown in Fig. 2B and Fig. 3C, indicates a spatial variation of $B_S$. As seen in Fig. 3C, the values of $B_S$, which are extracted from the values of $\Delta_{-1}$, range from 0.45 T to 0.75 T for positions 2-13, indicating a relatively uniform pseudo-magnetic field in this particular geometry. This is also in good agreement with the result obtained from the pseudo-Landau levels measurement in zero magnetic field (Fig. S2).

The combination of the pseudo-magnetic field and the external magnetic field in graphene monolayer leads to unconventional valley-polarized Landau quantization not present in graphene in either field. We envision probing more exotic phenomena if the pseudo-magnetic field and the external magnetic field are combined with other properties of graphene monolayer and the additional degrees of freedom found in graphene bilayers and multilayers.

**Acknowledgments:** This work was supported by the National Basic Research Program of China (Grants Nos. 2014CB920903, 2013CBA01603), the National Natural Science Foundation of China (Grant Nos. 11422430, 11374035), the program for New Century Excellent Talents in University of the Ministry of Education of China (Grant No. NCET-13-0054), Beijing Higher Education Young Elite Teacher Project (Grant No. YETP0238). We thank Zhongfan Liu and Mengxi Liu for the help in the synthesis of the sample.

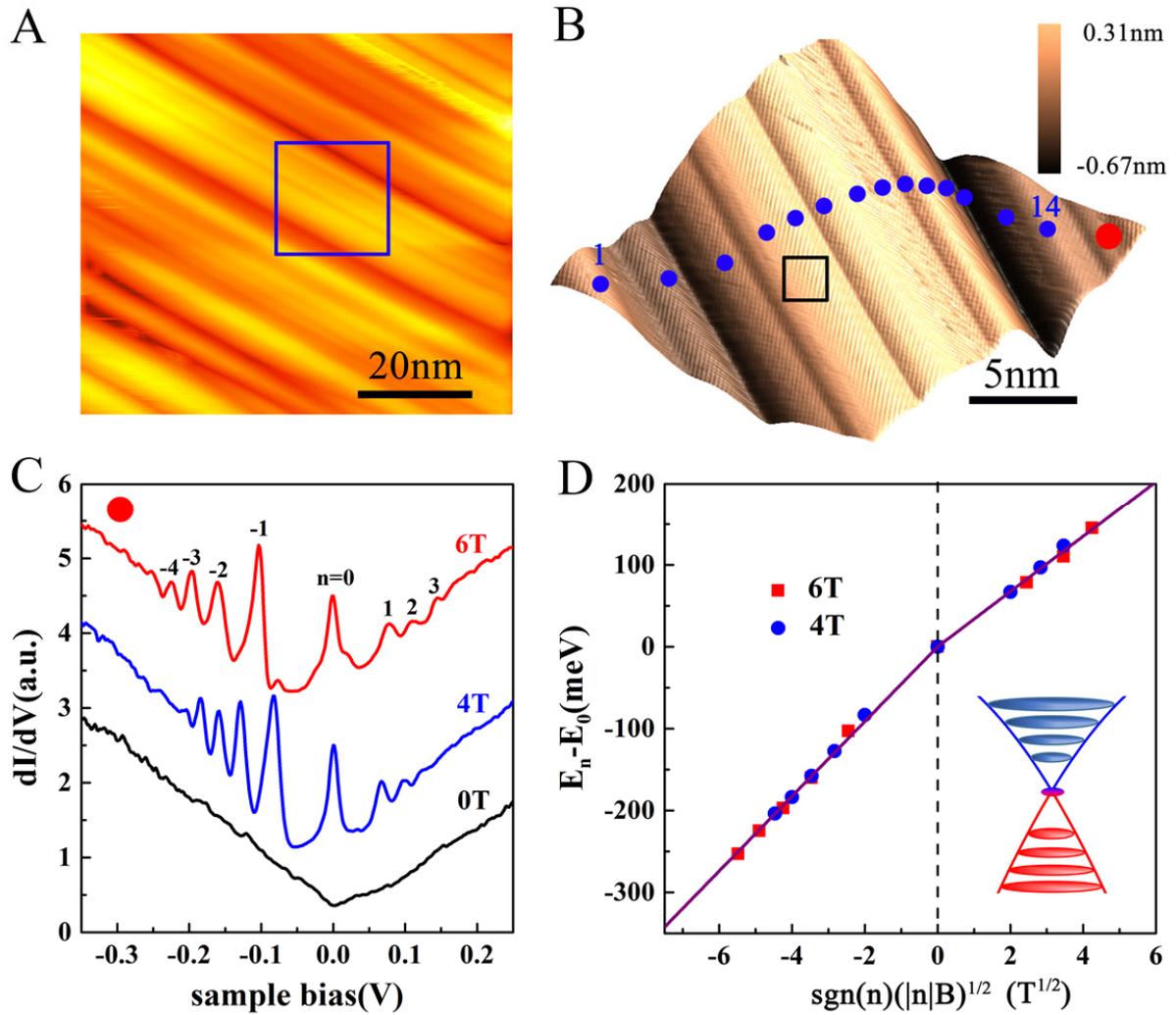

**Fig. 1. STM images and STS spectra of strained graphene monolayer on Rh foil.** (**A**) A representative STM image of a strained graphene region showing one-dimensional quasi-periodic ripples on Rh foil ($V_{sample}$ = -880 mV and $I$ = 0.4 nA). (**B**) An enlarged STM image of the graphene ripples in the blue frame in panel (**A**). (**C**) STS spectra, i.e., $dI/dV$-$V$ curves, taken at the position marked with red solid circle in panel (**B**) under different magnetic fields. For clarity, the curves are offset on the Y-axis and LL indices of massless Dirac fermions are marked. The spectra are shifted to make the $n$ = 0 LL stay at the same bias. (**D**) LL peak energies for different magnetic fields taken from panel (**C**) show a linear dependence against $sgn(n)(|n|B)^{1/2}$,

as expected for massless Dirac fermions in the graphene monolayer. The solid lines are linear fits of the data with Eq. (1), yielding the Fermi velocities $(1.257 \pm 0.009) \times 10^6$ m/s and $(0.930 \pm 0.014) \times 10^6$ m/s, for electrons and holes, respectively. Inset: Schematic of LLs in monolayer graphene in the quantum Hall regime with considering the next nearest-neighbor hopping energy $t'$.

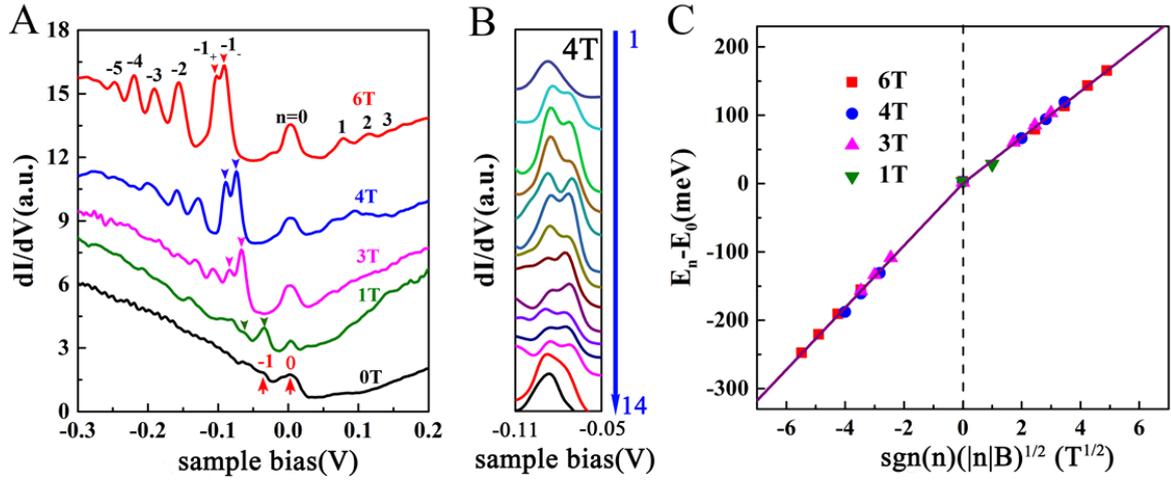

**Fig. 2. Pseudo-Landau levels and valley-polarized Landau quantization probed by STM. (A)** STS spectra taken in the area marked with black frame in Fig.1B under different magnetic fields. For clarity, the curves are offset on the Y-axis and LL indices of massless Dirac fermions are marked. For the case of $B = 0$ T, strained-induced pseudo-Landau levels, as marked by the red numbers, are observed. In the presence of external magnetic field, the $n = -1$ LL splits into two peaks, $-1_-$ and $-1_+$ (here -/+ denotes $K$ and $K'$ valleys respectively), and the energy spacing of the two peaks increases with decreasing the magnetic fields. **(B)** Several STS spectra around the $n = -1$ LL measured in the magnetic field of 4 T at different positions (blue dots in Fig. 1B). The energy difference of the two peaks of $n = -1$ LL, $\Delta_{-1}$, depends on the recorded positions. **(C)** LL peak energies for different magnetic fields taken from panel (**A**) show a linear dependence against $\mathrm{sgn}(n)(|n|B)^{1/2}$. For clarity, the data of the split $n = -1$ LL are not plotted in this figure. The large electron-hole asymmetry, which is similar to that shown in Fig. 1D, is also observed. The Fermi velocities for electrons and holes are measured to be $(1.250 \pm 0.007) \times 10^6$ m/s and $(0.927 \pm 0.007) \times 10^6$ m/s, respectively.

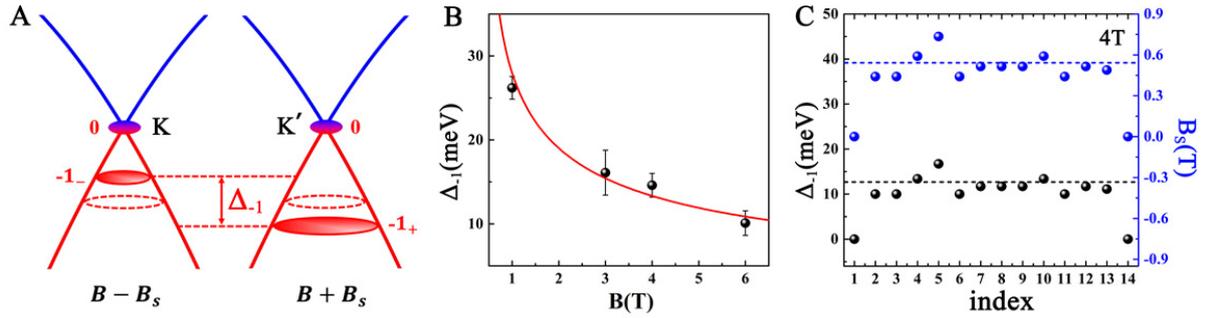

**Fig. 3. Valley-polarized Landau level and pseudo-magnetic fields in the strained graphene region. (A)** A schematic image showing Landau quantization in graphene monolayer in the presence of both the pseudo-magnetic fields and the external magnetic fields. The valley degeneracy of the $n = -1$ LL is lifted and the energy spacing of the valley-polarized LLs is described by Eq. (2). **(B)** This figure summarizes the $\Delta_{-1}$ measured at the same position as a function of the external magnetic fields. The solid curve is a fit to Eq. (2) with the only fitting parameter $B_S = (0.59 \pm 0.03)$ T. **(C)** This figure summarizes the $\Delta_{-1}$ measured in the magnetic field of 4 T at different positions (blue dots in Fig. 1B). The values of $B_S$, which are extracted from the values of $\Delta_{-1}$, are also plotted as a function of positions.

**Supplementary Materials:**

Materials and Methods

Figures S1-S3

References (*S1-S6*)